\documentclass[12pt]{article}

\usepackage{amsmath}
\usepackage{amsfonts}
\usepackage{amscd}
\usepackage{amsthm}
\usepackage{setspace}
\usepackage{graphicx}
\usepackage{authblk}
\usepackage{caption}
\usepackage{ytableau}
\usepackage{mathtools}

\def\Z{\mathbb{Z}}

\def\R{\mathbb{R}}
\def\C{\mathbb{C}}
\def\P{\mathbb{P}}

\def\til{\tilde}

\begin{document}

\begin{titlepage}

\begin{flushright}
YITP-16-35
\end{flushright}

\vskip 1cm

\begin{center}

{\large Gauge symmetries and matter fields in F-theory models without section\\
- compactifications on double cover and Fermat quartic K3 constructions times K3}

\vskip 1.2cm

Yusuke Kimura$^1$
\vskip 0.4cm
{\it $^1$Yukawa Institute for Theoretical Physics, Kyoto University, Kyoto 606-8502, Japan}
\vskip 0.4cm
E-mail: kimura@yukawa.kyoto-u.ac.jp

\vskip 1.5cm
\abstract{We investigate gauge theories and matter fields in F-theory compactifications on genus-one fibered Calabi--Yau 4-folds without a global section. In this study, genus-one fibered Calabi--Yau 4-folds are built as direct products of a genus-one fibered K3 surface that lacks a section times a K3 surface. We consider i) double covers of $\P^1\times\P^1$ ramified along a bidegree (4,4) curve, and ii) complete intersections of two bidegree (1,2) hypersurfaces in $\P^1\times\P^3$ to construct genus-one fibered K3 surfaces without a section. $E_7$ gauge group arises in some F-theory compactifications on double covers times K3. We show that the tadpole can be cancelled for an F-theory compactification on complete intersection K3 times K3, when complete intersection K3 is isomorphic to the Fermat quartic, and the complex structure of the other K3 surface in the direct product is appropriately chosen.}  

\end{center}
\end{titlepage}

\tableofcontents
\section{Introduction}
F-theory \cite{Vaf, MV1, MV2} is a nonperturbative extension of type IIB superstring theory. F-theory is compactified on Calabi--Yau manifolds with a torus fibration. In F-theory compactification, 7-branes are wrapped on the irreducible components of the discriminant locus in the base space. The discriminant locus is the codimension one locus in the base, along which torus fibers degenerate. 
\par In F-theory, non-Abelian gauge symmetry on 7-branes is determined by the type of singular fibers over a discriminant component, on which the 7-branes are wrapped. Matter representations arise from rank one enhancements of singularities of a compactification space \cite{BIKMSV, KV, GM, MTmatter, GM2}. See \cite{KMP, Pha} for discussion of other types of matter representations that arise from the structure of divisor. The resolution and deformation of singularities were discussed in \cite{KM}. \cite{DWmodel, BHV1} analyzed matter in four-dimensional F-theory in the presence of a flux. 
\par F-theory compactifications on Calabi--Yau elliptic fibrations that admit a global section have been discussed, for example, in \cite{DW, CGH, GW, KMSS, GKPW, MP, MPW, BGK, BMPWsection, CKP, CGKP, BMPW, SNW}. There are Calabi--Yau manifolds with a torus fibration that do not have a global section. Recently, initiated in \cite{BM, MTsection}, F-theory compactifications on genus-one fibered Calabi--Yau manifolds that lack a global section\footnote{\cite{BEFNQ, BDHKMMS} considered F-theory on genus-one fibrations without a global section.} have been investigated in several studies. See also, e.g., \cite{AGGK, KMOPR, GGK, MPTW, MPTW2, CDKPP, LMTW, K} for discussion of F-theory compactifications on genus-one fibrations lacking a global section.
\par In this note, we construct genus-one fibered Calabi--Yau 4-folds without a global section, and we investigate gauge theories and matter fields in F-theory compactifications on these spaces. We build genus-one fibered Calabi--Yau 4-folds as direct products of K3 surfaces, K3$\times$K3. 
\par We consider two constructions of genus-one fibered K3 surfaces: \\
1) double covers of $\P^1\times\P^1$ ramified along a bidegree (4,4) curve and \\
2)complete intersections of two bidegree (1,2) hypersurfaces in $\P^1\times\P^3$ \\ 
Generic members of these families do not have a global section to the fibration. The direct product of such genus-one fibered K3 surface without a section and a K3 surface gives a genus-one fibered Calabi--Yau 4-fold without a section. 
\par Among K3 surfaces in families 1) and 2), we focus on the members given by specific forms of equations, to perform detailed analysis of gauge theories and matter fields that arise in F-theory compactifications. For the family of double covers 1), we particularly consider the K3 surfaces whose genus-one fibers possess complex multiplication of order 4. For the family of complete intersections 2), we particularly consider the members that are isomorphic to the Fermat quartic. We will see in Section \ref{ssec:3.1} that $E_7$ gauge group arises on 7-branes in F-theory compactifications on special double covers times K3. 

\par The outline of this note is as follows: In Section \ref{sec:2}, we introduce families of genus-one fibered K3 surfaces that lack a global section. In Section \ref{sec:3}, we deduce non-Abelian gauge groups arising on the 7-branes in F-theory compactifications on these K3 surfaces times K3. For double covers whose genus-one fibers have complex multiplication of order 4, we perform a consistency check of the gauge symmetries, by considering the anomaly cancellation condition and the possible monodromies around the singular fibers. Those fibers that possess complex multiplication of order 4 impose strong constraints on the possible monodromies around the singular fibers; this observation limits allowed gauge symmetries on the 7-branes. Similar consistency check of gauge symmetries using the anomaly cancellation condition and the allowed monodromies around the singular fibers can be found in \cite{K} \footnote{\cite{K} concerns genus-one fibered K3 surfaces whose fibers have complex multiplication of order 3.}. For complete intersections that are isomorphic to the Fermat quartic, we confirm that the non-Abelian gauge symmetries on the 7-branes satisfy the anomaly cancellation condition. We also determine the Jacobian fibrations of double covers and complete intersections. As a result, we find that F-theory compactifications on some K3 genus-one fibrations without a global section times K3 do not have a $U(1)$ gauge field. In Section \ref{sec:4}, we compute potential matter fields that arise on the 7-branes in F-theory compactifications on constructed K3 genus-one fibrations times K3. We consider an F-theory compactification with a 4-form flux \cite{BB, SVW, W, GVW, DRS} turned on. Including flux breaks half of $N=2$ supersymmetry in F-theory on K3 $\times$ K3. In this flux compactification, hypermultiplets in four-dimensional $N=2$ theory split into vector-like pairs. We find that the tadpole can be cancelled for F-theory flux compactifications on the Fermat quartic times some appropriate attractive K3 surface. Therefore, we confirm that vector-like pairs in fact arise for this particular case. We state concluding remarks in Section \ref{sec:5}.

\section{Two families of K3 surfaces without section}
\label{sec:2}
In this section, we introduce three families of genus-one fibered K3 surfaces that do not admit a global section. 

\subsection{Double covers of $\P^1\times\P^1$ ramified along bidegree (4,4) Curve}
\label{ssec 2.1}
Double covers of $\P^1\times\P^1$ ramified over a bidegree (4,4) curve have the trivial canonical bundle; therefore, these surfaces are K3 surfaces. A fiber of a projection onto $\P^1$ is a double cover of $\P^1$ ramified over 4 points, which is a genus-one curve. Therefore, projection onto $\P^1$ gives a genus-one fibration. 
\par We show that generic members of double covers of $\P^1\times\P^1$ ramified over a bidegree (4,4) curve do not admit a section to the fibration. Let $p_1$ and $p_2$ denote the projections from $\P^1\times \P^1$ onto the first $\P^1$ in the product $\P^1\times\P^1$ and onto the second $\P^1$, respectively: 
$$
\begin{CD}
\P^1\times\P^1 @>{p_1}>> \P^1 \\
@V{p_2}VV \\
\P^1.
\end{CD}
$$ 
Let $\mathcal{O}_{\P^1}(1)$ denote a point class in $\P^1$. The pullback of a point class in $\P^1$ under the projection $p_1$ is $p^*_1\mathcal{O}_{\P^1}(1)=\{{\rm pt}\}\times \P^1$, and the pullback of a point class in $\P^1$ under $p_2$ is $p^*_2\mathcal{O}_{\P^1}(1)=\P^1\times\{{\rm pt}\}$. Therefore, $(p^*_1\mathcal{O}_{\P^1}(1))^2=0$, $(p^*_2\mathcal{O}_{\P^1}(1))^2=0$, and $p^*_1\mathcal{O}_{\P^1}(1)\cdot p^*_2\mathcal{O}_{\P^1}(1)=1$. 
\par Let $S$ denote a double cover of $\P^1\times\P^1$ ramified over a curve of bidegree (4,4). The projections $p_1$ and $p_2$ induce projections $\til{p}_1$ and $\til{p_2}$ from $S$ onto the $\P^1$'s. Each of the projection $\til{p}_1$ and $\til{p}_2$ gives a genus-one fibration. 
\par Let $D_1:=\til{p}^*_1\mathcal{O}_{\P^1}(1)$ and $D_2:=\til{p}^*_2\mathcal{O}_{\P^1}(1)$ be the pullbacks of a point in $\P^1$ to $S$ under the projections $\til{p}_1$ and $\til{p}_2$. Since $S$ is a double cover of $\P^1\times\P^1$, the intersection numbers of the pullbacks to $S$ are twice those in $\P^1\times\P^1$: $D_1^2=0$, $D_2^2=0$, and $D_1\cdot D_2=2$. The generic N\'eron--Severi lattice of double cover $S$ is generated by $D_1$ and $D_2$ \cite{Moi}, and therefore it has the intersection matrix 
\begin{equation}
\begin{pmatrix}
0 & 2 \\
2 & 0 \\
\end{pmatrix}. 
\end{equation} The divisor $D_1$ has self-intersection $0$; therefore, it represents the fiber class $F$. $D_2$ represents a 2-section. Every divisor has the intersection number that is a multiple of 2 with the fiber $F$; thus, the double cover $S$ does not have a global section.   
\par The equation of a double cover of $\P^1\times\P^1$ ramified over a bidegree (4,4) curve is given by:
\begin{equation}
\tau^2=b_1(t)x^4+b_2(t)x^3+b_3x^2+b_4(t)x+b_5(t),
\label{eq:double cover 1}
\end{equation}
where $x$ is the inhomogeneous coordinate on the first $\P^1$ in the product $\P^1\times\P^1$ and $t$ is the inhomogeneous coordinate on the second $\P^1$. $t$ is the coordinate on the base $\P^1$.
\par To study non-Abelian gauge groups and matter fields on the 7-branes in F-theory compactifications, in this note, we focus on the double covers of $\P^1\times\P^1$ given by the equations of the following form:
\begin{equation}
\tau^2=a_1(t)x^4+a_2(t),
\label{eq:2}
\end{equation}
where $a_1(t)$ and $a_2(t)$ are polynomials in $t$ of the highest degree of 4. 
\par Equation (\ref{eq:2}) has the automorphism group $\Z_4$ generated by the map
\begin{equation}
x\rightarrow e^{\rm 2\pi i/4}x.
\end{equation}
From this, we can see that genus-one fibers of the double covers given by equation (\ref{eq:2}) possess complex multiplication of order 4. It is known that a genus-one curve possessing complex multiplication of order 4 has j-invariant 1728. Therefore, the complex structure of smooth genus-one fibers of a double cover given by equation (\ref{eq:2}) is constant over the base, specified by j-invariant 1728. This forces the singular fibers of double cover (\ref{eq:2}) to have j-invariant 1728. This greatly constrains the possible gauge symmetries arising on the 7-branes in F-theory compactifications on double covers (\ref{eq:2}) times K3.
\par We use this property in Section \ref{ssec:3.4} to perform a consistency check of the non-Abelian gauge symmetries on the 7-branes, which will be deduced in Section \ref{ssec:3.1}. 

\subsection{Complete intersections of two bidegree (1,2) hypersurfaces in $\P^1\times\P^3$}
\label{ssec:2.2}
Complete intersections of two bidegree (1,2) hypersurfaces in $\P^1\times\P^3$ are K3 surfaces. Projection onto $\P^1$ is a complete intersection of two degree 2 curves in $\P^3$, which is a genus-one curve. Therefore, projection onto $\P^1$ gives a genus-one fibration; (1,2) and (1,2) complete intersection in $\P^1\times\P^3$ is a genus-one fibered K3 surface. 
\par $\P^1\times\P^3$ has the projections $q_1$ and $q_2$ onto $\P^1$ and $\P^3$, respectively:
$$
\begin{CD}
\P^1\times\P^3 @>{q_2}>> \P^3 \\
@V{q_1}VV \\
\P^1.
\end{CD}
$$ 
Let $\til{q_1}$ and $\til{q_2}$ be the restrictions of the projections $q_1$ and $q_2$ to (1,2) and (1,2) complete intersection K3 in $\P^1\times\P^3$. For notational simplicity, we set $D_3:=\til{q}^*_1\mathcal{O}_{\P^1}(1)$ and $D_4:=\til{q}^*_2\mathcal{O}_{\P^3}(1)$. Then, we have $D^2_3=0$; therefore, $D_3$ represents the fiber class $F$. $D^2_4$ is the intersection number of two bidegree (1,2) curves in $\P^1\times\P^1$, which is 4. Finally, $D_3\cdot D_4$ is the intersection number of two conics in $\P^2$, which is 4. Therefore, the intersection matrix of the N\'eron--Severi lattice of generic (1,2) and (1,2) complete intersection K3 in $\P^1\times\P^3$ is 
\begin{equation}
\begin{pmatrix}
0 & 4 \\
4 & 4 \\
\end{pmatrix}.
\label{eq:Gram}
\end{equation}
The generators of intersection matrix (\ref{eq:Gram}) represent the fiber class and a 4-section. Any divisor has an intersection number that is a multiple of 4 with the fiber class $F$; therefore, a generic member of (1,2) and (1,2) complete intersections in $\P^1\times\P^3$ does not admit a global section.
\par Note that the natural projection $\til{q_2}$ is an isomorphism from (1,2) and (1,2) complete intersection K3 onto the image in $\P^3$. This can be seen as follows: let $[t_0:t_1]$ be the homogeneous coordinates on $\P^1$. Then, a complete intersection of bidegree (1,2) and (1,2) hypersurfaces in $\P^1\times\P^3$ is given by simultaneous vanishing of the following two equations: 
\begin{eqnarray}
\label{eq:predet}
f_1t_0+f_2t_1 & = & 0\\ 
g_1t_0+g_2t_1 & = & 0 \nonumber
\end{eqnarray} 
where $f_i$ and $g_i$ ($i=1,2$) are polynomials on $\P^3$ of degrees of 2. Next, we consider the image of the projection $\til{q_2}$ of complete intersection K3 (\ref{eq:predet}) into $\P^3$. The equation of the image of complete intersection K3 (\ref{eq:predet}) in $\P^3$ does not depend on the coordinates $[t_0:t_1]$. Therefore, the defining equation of the projection image of complete intersection K3 (\ref{eq:predet}) in $\P^3$ is given by vanishing of the determinant of the equation (\ref{eq:predet}):
\begin{equation}
{\rm det}\begin{pmatrix}
f_1 & f_2 \\
g_1 & g_2 \\
\end{pmatrix}=0.
\label{eq:determinantal}
\end{equation} Thus, the projection image of complete intersection K3 (\ref{eq:predet}) in $\P^3$ is the determinantal locus given by equation (\ref{eq:determinantal}). Equation (\ref{eq:determinantal}) is quartic, and therefore the projection image is a degree 4 hypersurface in $\P^3$, which is a K3 surface. The projection $\til{q_2}$ from complete intersection K3 (\ref{eq:predet}) to K3 (\ref{eq:determinantal}) in $\P^3$ gives a morphism between the K3 surfaces. The inverse image of a point is a point under this morphism, and the K3 surface is minimal; therefore, this morphism is an isomorphism.  
\par In this study, we focus on the (1,2) and (1,2) complete intersection in $\P^1\times\P^3$ given by the following equation:
\begin{eqnarray}
\label{array 1}
x^2_1+x^2_3+2tx_2x_4 & = & 0 \\
x^2_2+x^2_4+2tx_1x_3 & = & 0 \nonumber
\end{eqnarray}
$[x_1:x_2:x_3:x_4]$ is the homogeneous coordinates on $\P^3$ and $t$ is the inhomogeneous coordinate on $\P^1$. We set $t:=t_1/t_0$ in equation (\ref{array 1}).
\par As the following argument shows, the complete intersection given by (\ref{array 1}) is isomorphic to the Fermat quartic: The projection image of complete intersection (\ref{array 1}) into $\P^3$ is the determinantal variety given by 
\begin{equation}
{\rm det}\begin{pmatrix}
x^2_1+x^2_3 & 2x_2x_4 \\
x^2_2+x^2_4 & 2x_1x_3 \\
\end{pmatrix}=0,
\end{equation}  
or equivalently, 
\begin{equation}
(x^2_1+x^2_3)x_1x_3=(x^2_2+x^2_4)x_2x_4.
\label{det var 1}
\end{equation}
The left-hand side of (\ref{det var 1}) is $(x_1+x_3)^4-(x_1-x_3)^4$ (times a constant), and the right-hand side equals $(x_2+x_4)^4-(x_2-x_4)^4$. The linear change of variables:
\begin{eqnarray}
\label{transformation Fermat}
x_1+x_3 & = & x \\ \nonumber
x_1-x_3 & = & e^{2\pi i/8}y \\ \nonumber
x_2+x_4 & = & e^{2\pi i/8}z \\ \nonumber
x_2-x_4 & = & w \nonumber
\end{eqnarray} 
transforms (\ref{det var 1}) into 
\begin{equation}
x^4+y^4+z^4+w^4=0 \subset \P^3.
\label{Fermat Quartic in 2.2}
\end{equation}
Surface (\ref{Fermat Quartic in 2.2}) is known as the {\it Fermat quartic}. Therefore, complete intersection K3 (\ref{array 1}) is isomorphic to the Fermat quartic surface (\ref{Fermat Quartic in 2.2}). Fermat quartic is known to be an attractive K3 surface\footnote{It is standard to call a K3 having the largest Picard number $\rho=20$ a singular K3 in mathematics. We call such a K3 an attractive K3 in this study, following the convention of the term used in \cite{M}.}, whose transcendental lattice $T_S$ has the intersection matrix $\begin{pmatrix}
8 & 0 \\
0 & 8 \\
\end{pmatrix}$.

\section{Gauge groups on 7-branes}
\label{sec:3}
In this section, we determine non-Abelian gauge symmetries that arise in F-theory compactifications on double covers of $\P^1\times\P^1$ ramified over a bidegree (4,4) curve times K3, and compactifications on (1,2) and (1,2) complete intersections in $\P^1\times\P^3$ times K3. We also check the consistency of the solutions by considering the monodromies around the singular fibers, and the anomaly cancellation condition. 
\par Generic fibers of a genus-one fibered surface\footnote{Elliptic surfaces and singular fibers are discussed in \cite{Kod, Ner, Shiodamodular, Tate, Shioda, Silv, BHPV, SchShio}. See \cite{Cas} for discussion of elliptic curves and the Jacobian fibration.} are smooth genus-one curves. Fibers degenerate to the singular fibers along the codimension 1 locus, which is called the discriminant locus, in the base. Kodaira \cite{Kod} classified the types of the singular fibers of a genus-one fibered surface. In this note, we use Kodaira's notations for the singular fiber types.
\par
\begin{figure}
\begin{center}
\includegraphics[width=\linewidth,height=\textheight,keepaspectratio]{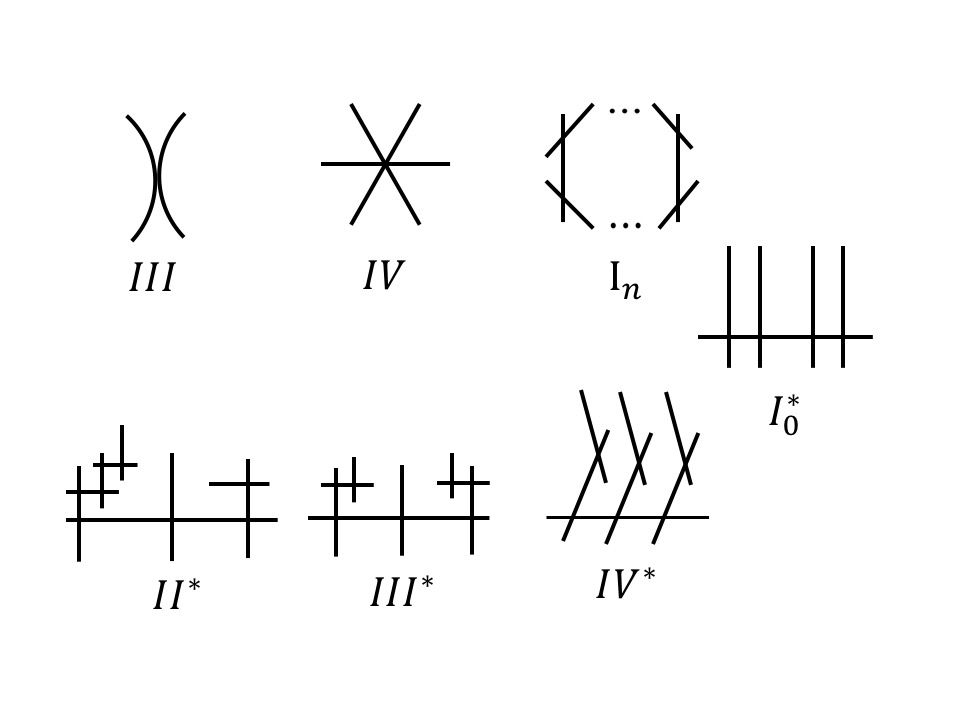}
\caption{\label{fig}Images of fiber types.}
\end{center}
\end{figure}

 A singular fiber of a genus-one fibered surface is either the sum of smooth $\P^1$'s intersecting in specific ways or a $\P^1$ with a single singularity. The latter case is not a singularity of a surface. Each of type $I_1$ fiber and type $II$ fiber is $\P^1$ with a single singularity: type $I_1$ fiber is a rational curve with a node and type $II$ fiber is a rational curve with a cusp. There are seven types of singular fibers that are reducible into the sum of smooth $\P^1$'s. They are two infinite series $I_n$ ($n\ge 2$) and $I^*_m$ ($m\ge 0$) and five types $III, IV, II^*, III^*$ and $IV^*$. See Figure \ref{fig} for images of singular fibers. Each line in an image represents a $\P^1$ component. The images in Figure \ref{fig} show the configurations of $\P^1$ components of the fiber types.
\par Non-Abelian gauge symmetries that arise on the 7-branes are determined by the types of the singular fibers. For discussion of the correspondence between the gauge groups on the 7-branes and the types of the singular fibers, see \cite{MV2, BIKMSV}. Table \ref{tabcorresp} below shows the correspondence between the fiber types and the singularity types for F-theory compactification. 

\begingroup
\renewcommand{\arraystretch}{1.5}
\begin{table}[htb]
\begin{center}
  \begin{tabular}{|c|c|} \hline
Fiber type & Singularity \\ \hline
$I_n$ & $A_{n-1}$ \\
$I^*_n$ & $D_{n+4}$ \\ 
$III$ & $A_1$ \\
$IV$ & $A_2$ \\ 
$IV^*$ & $E_6$ \\
$III^*$ & $E_7$ \\
$II^*$ & $E_8$ \\ \hline   
\end{tabular}
\caption{\label{tabcorresp}Fiber type and singularity type correspondence.}
\end{center}
\end{table}  
\endgroup 

\subsection{Non-Abelian gauge groups on double covers of $\P^1\times \P^1$ ramified over a bidegree (4,4) curve times K3}
\label{ssec:3.1}
As discussed in Section \ref{ssec 2.1}, in this note, we focus on double covers of $\P^1\times\P^1$ ramified over a bidegree (4,4) curve, given by the following specific form of equations:
\begin{equation}
\tau^2=a_1(t)x^4+a_2(t).
\label{double cover no expansion in 3.1}
\end{equation}
$a_1(t)$ and $a_2(t)$ are degree 4 polynomials in the variable $t$. By splitting $a_1(t)$ and $a_2(t)$ into linear factors, (\ref{double cover no expansion in 3.1}) may be rewritten as the following equation:
\begin{equation}
\tau^2=\Pi^4_{i=1}(t-\alpha_i)\, x^4+\Pi^8_{j=5}(t-\alpha_j).
\label{expand double cover in 3.1}
\end{equation}
The Jacobian fibration of double cover (\ref{expand double cover in 3.1}) is given by \cite{Muk}:
\begin{equation}
\tau^2=\frac{1}{4}x^3-\Pi^8_{i=1}(t-\alpha_i)x.
\label{jacobian in 3.1}
\end{equation}
The discriminant of Jacobian (\ref{jacobian in 3.1}) is given by
\begin{equation}
\Delta \sim \Pi^8_{i=1}(t-\alpha_i)^3.
\label{discriminant in 3.1}
\end{equation}
\par (\ref{jacobian in 3.1}) is the Weierstrass form, therefore we can determine the types of the singular fibers from the vanishing orders of the coefficient of the equation (\ref{jacobian in 3.1}) and the discriminant (\ref{discriminant in 3.1}). The correspondence of the fiber types and the vanishing orders of the coefficients of the Weierstrass form is shown in Table \ref{tabweierstrass} below. 
\begingroup
\renewcommand{\arraystretch}{1.5}
\begin{table}[htb]
\begin{center}
  \begin{tabular}{|c|c|c|c|} \hline
Fiber type & Order of $f$ & Order of $g$ & Order of $\Delta$ \\ \hline
$I_0$ & $\ge 0$ & $\ge 0$ & 0 \\ \hline
$I_n$ ($n\ge 1$) & 0 & 0 & $n$ \\ \hline
$II$ & $\ge 1$ & 1 & 2 \\ \hline
$III$ & 1 & $\ge 2$ & 3 \\ \hline
$IV$ & $\ge 2$ & 2 & 4 \\ \hline
$I^*_0$ & $\ge 2$ & 3 & 6 \\ \cline{2-4}
 & 2 & $\ge 3$ & 6 \\ \hline
$I^*_n$ ($n \ge 1$) & 2 & 3 & $n+6$ \\ \hline
$IV^*$ & $\ge 3$ & 4 & 8 \\ \hline
$III^*$ & 3 & $\ge 5$ & 9 \\ \hline
$II^*$ & $\ge 4$ & 5 & 10 \\ \hline   
\end{tabular}
\caption{\label{tabweierstrass}Vanishing orders of the coefficients of the Weierstrass form $y^2=x^3+fx+g$ and the discriminant $\Delta$, and the corresponding fiber types.}
\end{center}
\end{table}  
\endgroup 
Since double cover (\ref{expand double cover in 3.1}) and Jacobian fibration (\ref{jacobian in 3.1}) have identical types of singular fibers over the same locations in the base, the result of singular fibers for Jacobian (\ref{jacobian in 3.1}) gives identical singular fibers of double cover (\ref{expand double cover in 3.1}). 
\par Singular fibers of Jacobian (\ref{jacobian in 3.1}) are at $t=\alpha_i$, $i=1,\cdots, 8$. When $\alpha_i$'s are mutually distinct, (i.e., $\alpha_i\ne\alpha_j$ when $i\ne j$), from Table \ref{tabweierstrass}, we find that the fiber type at $t=\alpha_i$, $i=1,\cdots, 8$, is $III$. Therefore, the non-Abelian gauge group that arises on the 7-branes in F-theory compactification on double cover (\ref{expand double cover in 3.1}) times K3 is
\begin{equation}
SU(2)^8.
\end{equation} 
\par From Table \ref{tabweierstrass}, we find that when the multiplicity of $\alpha_i$ is 2, i.e., when there is $j$, $j\ne i$, such that $\alpha_i=\alpha_j$, the fiber type at $t=\alpha_i$ is $I^*_0$. Therefore, we deduce that when two type $III$ fibers collide, $SU(2)^2$ gauge group on the 7-branes is enhanced to $SO(8)$ gauge group. When the multiplicity of $\alpha_i$ is 3, the fiber type at $t=\alpha_i$ is $III^*$. This means that, when a triplet of type $III$ fibers coincide, $SU(2)^3$ gauge group on the 7-branes is enhanced to $E_7$ gauge group. When the multiplicity of $\alpha_i$ becomes greater than 3, the Calabi--Yau condition is broken. Thus, we find that the most enhanced gauge symmetry on the 7-branes is
\begin{equation}
E_7\times E_7\times SO(8). 
\end{equation}
\par When the gauge group on the 7-branes is $E_7\times E_7\times SO(8)$, double cover of $\P^1\times\P^1$ (\ref{double cover no expansion in 3.1}) becomes an attractive K3 surface. As we will see in Section \ref{ssec:3.5}, the Jacobian of this attractive K3 surface has the Mordell--Weil group of rank 0. Therefore, when the non-Abelian gauge group on the 7-branes is $E_7\times E_7\times SO(8)$, the gauge group in an F-theory compactification on double cover of $\P^1\times \P^1$ (\ref{double cover no expansion in 3.1}) times K3 does not have a $U(1)$ gauge field.
\par To be explicit, we consider an example, in which the gauge group on the 7-branes becomes $E_7\times E_7\times SO(8)$. We consider the double cover of $\P^1\times\P^1$ given by the equation:
\begin{equation}
\tau^2=(t-\alpha_1)^3(t-\alpha_2)x^4+(t-\alpha_2)(t-\alpha_3)^3.
\label{example 1 in 3.1}
\end{equation} 
The singular fibers are at $t=\alpha_1,\alpha_2,\alpha_3$. ($\alpha_1, \alpha_2, \alpha_3$ are mutually distinct.) Fiber type is $III^*$ at $t=\alpha_1,\alpha_3$, and fiber type is $I^*_0$ at $t=\alpha_2$. The gauge group on the 7-branes is $E_7\times E_7\times SO(8)$ in the F-theory compactification on double cover (\ref{example 1 in 3.1}) times K3.

\subsection{Non-Abelian gauge groups on Fermat quartic times K3}
\label{ssec:3.2}
We saw in Section \ref{ssec:2.2} that the complete intersection of two (1,2) hypersurfaces in $\P^1\times\P^3$ given by
\begin{eqnarray}
\label{complete intersection in 3.2}
x^2_1+x^2_3+2tx_2x_4 & = & 0 \\
x^2_2+x^2_4+2tx_1x_3 & = & 0 \nonumber
\end{eqnarray}
($[x_1:x_2:x_3:x_4]$ are the homogeneous coordinates on $\P^3$ and $t$ is the inhomogeneous coordinate on $\P^1$) is isomorphic to the Fermat quartic
\begin{equation}
x^4+y^4+z^4+w^4=0 \subset \P^3.
\label{Fermat Quartic in 3.2}
\end{equation}
In this section, we determine the non-Abelian gauge symmetry that arises in the F-theory compactification on Fermat quartic (\ref{complete intersection in 3.2}) times K3. 
\par We compute the Jacobian fibration of complete intersection (\ref{complete intersection in 3.2}) to determine the singular fibers of the complete intersection (\ref{complete intersection in 3.2}). We introduce a parameter $\lambda$ and add $-\lambda$ times the second equation to the first equation in (\ref{complete intersection in 3.2}) to obtain: 
\begin{equation}
x^2_1+x^2_3+2tx_2x_4-\lambda(x^2_2+x^2_4+2tx_1x_3).
\label{subtract 3.2}
\end{equation}
We arrange the coefficients of (\ref{subtract 3.2}) into a symmetric matrix: 
\begin{equation}
\begin{pmatrix}
1 & 0 & -t\lambda & 0 \\
0 & -\lambda & 0 & t \\
-t\lambda & 0 & 1 & 0 \\
0 & t & 0 & -\lambda 
\end{pmatrix}.
\label{matrix 3.2}
\end{equation}
We compute the determinant of 4 $\times$ 4 matrix (\ref{matrix 3.2}) to obtain the equation of the Jacobian fibration of complete intersection (\ref{complete intersection in 3.2}): 
\begin{equation}
\tau^2=-t^2\lambda^4+(t^4+1)\lambda^2-t^2.
\label{jacobian Fermat in 3.2}
\end{equation} 
Jacobian (\ref{jacobian Fermat in 3.2}) is a double cover of $\P^1\times\P^1$; $\lambda$ and $t$ are the inhomogeneous coordinates on the first $\P^1$ and second $\P^1$ in the product $\P^1\times\P^1$, respectively. 
\par Complete intersection K3 (\ref{complete intersection in 3.2}) and Jacobian (\ref{jacobian Fermat in 3.2}) have identical discriminant loci and singular fiber types. Thus, we can determine the types and locations of the singular fibers of complete intersection K3 (\ref{complete intersection in 3.2}) by computing the singular fibers of Jacobian (\ref{jacobian Fermat in 3.2}). Jacobian (\ref{jacobian Fermat in 3.2}) transforms into the following extended Weierstrass form:
\begin{equation}
\label{extended form in 3.2}
y^2=\frac{1}{4}x^3-\frac{1}{2}(t^4+1)x^2+\frac{1}{4}(t^4-1)^2x.
\end{equation}
The discriminant of Jacobian (\ref{jacobian Fermat in 3.2}) is
\begin{equation}
\Delta \sim 16t^4(t^4-1)^4.
\label{disc Fermat in 3.2}
\end{equation} 
Therefore, we find from the discriminant (\ref{disc Fermat in 3.2}) that the Jacobian fibration (\ref{jacobian Fermat in 3.2}) has six singular fibers at $t=0,\infty, \pm 1, \pm i$. By completing the cube, the extended Weierstrass form (\ref{extended form in 3.2}) can be transformed into the Weierstrass form:
\begin{equation}
y^2=\frac{1}{4}x^3-\frac{1}{12}(t^8+14t^4+1)x+\frac{1}{54}(t^{12}-33t^8-33t^4+1).
\label{resulting Weierstrass in 3.2}
\end{equation}
We study the coefficients of the resulting Weierstrass form (\ref{resulting Weierstrass in 3.2}) to find that the singular fibers are $I_n$ fibers for some $n$. Thus, by studying the orders of the zeros of the discriminant (\ref{disc Fermat in 3.2}), we conclude that complete intersection (\ref{complete intersection in 3.2}) has six $I_4$ fibers at $t=0,\infty, \pm 1, \pm i$. 
\par We deduce from the discussion above that the gauge group that arises on the 7-branes in the F-theory compactification on Fermat quartic (\ref{complete intersection in 3.2}) times K3 is
\begin{equation}
SU(4)^6. 
\end{equation}
We will see in Section \ref{ssec:3.5} that this F-theory compactification does not have a $U(1)$ gauge symmetry.
\par Fermat quartic is known to be an attractive K3, with a transcendental lattice\footnote{The complex structure of an attractive K3 is specified by its transcendental lattice \cite{SI}. See Section \ref{ssec:4.1} for the relationship of complex structure and the transcendental lattice.} 
\begin{equation}
T_S=\begin{pmatrix}
8 & 0 \\
0 & 8 \\
\end{pmatrix}.
\end{equation} Therefore, in this note, we denote the Fermat quartic by $S_{[8 \hspace{1mm} 0 \hspace{1mm} 8]}$. Six $I_4$ fibers have reducible fiber type $A^6_3$. From Table 2 of \cite{SZ}, we see that attractive K3 with reducible fiber type $A_3^6$ {\it with a section} is unique, and it has the transcendental lattice $T_S=\begin{pmatrix}
4 & 0 \\
0 & 4 \\
\end{pmatrix}$. Attractive K3 with $T_S=\begin{pmatrix}
4 & 0 \\
0 & 4 \\
\end{pmatrix}$ is not the Fermat quartic; therefore, we conclude that the Fermat quartic with six $I_4$ fibers does not have a global section.
\par Jacobian (\ref{jacobian Fermat in 3.2}) is attractive K3 with a section with six $I_4$ fibers; therefore, we deduce that Jacobian (\ref{jacobian Fermat in 3.2}) is attractive K3 with the transcendental lattice $T_S=\begin{pmatrix}
4 & 0 \\
0 & 4 \\
\end{pmatrix}$. We denote Jacobian (\ref{jacobian Fermat in 3.2}) by $S_{[4 \hspace{1mm} 0 \hspace{1mm} 4]}$\footnote{\cite{Sch} mentions the facts that the Fermat quartic with six $I_4$ fibers does not admit a section, and the Jacobian of the Fermat quartic $S_{[8 \hspace{1mm} 0 \hspace{1mm} 8]}$ with six $I_4$ fibers has the transcendental lattice $\begin{pmatrix}
4 & 0 \\
0 & 4 \\
\end{pmatrix}$.}.

\subsection{Monodromy and anomaly cancellation condition}
\label{ssec:3.3}
\subsubsection{Monodromies around singular fibers}
Each singular fiber type has the specific monodromy, which takes value in the special linear group $SL_2(\Z)$, and each fiber type has the specific j-invariant. Monodromies and their orders, the j-invariants of singular fibers, and the number of 7-branes\footnote{Euler numbers of the singular fiber types were computed in \cite{Kod}. Euler numbers of the fiber types have an interpretation as the numbers of 7-branes associated to the fiber types.} associated with fiber types are displayed in Table \ref{table monodromy in 3.3.1} below. Kodaira \cite{Kod} derived the results in Table \ref{table monodromy in 3.3.1}. ``Regular'' for the j-invariant of fiber type $I^*_0$ in Table \ref{table monodromy in 3.3.1} means that the j-invariant of fiber type $I^*_0$ may take any finite value in $\C$. The value of the j-invariant of fiber type $I^*_0$ depends on the situations. 

\begingroup
\renewcommand{\arraystretch}{1.1}
\begin{table}[htb]
  \begin{tabular}{|c|c|r|c|c|} \hline
Fiber type & J-invariant & Monodromy  & Order of Monodromy & \# of 7-branes (Euler number) \\ \hline
$I^*_0$ & regular & $-\begin{pmatrix}
1 & 0 \\
0 & 1 \\
\end{pmatrix}$ & 2 & 6\\ \hline
$I_b$ & $\infty$ & $\begin{pmatrix}
1 & b \\
0 & 1 \\
\end{pmatrix}$ & infinite & $b$\\
$I^*_b$ & $\infty$ & $-\begin{pmatrix}
1 & b \\
0 & 1 \\
\end{pmatrix}$ & infinite & $b+$6\\ \hline
$II$ & 0 & $\begin{pmatrix}
1 & 1 \\
-1 & 0 \\
\end{pmatrix}$ & 6 & 2\\
$II^*$ & 0 & $\begin{pmatrix}
0 & -1 \\
1 & 1 \\
\end{pmatrix}$ & 6 & 10\\ \hline
$III$ & 1728 & $\begin{pmatrix}
0 & 1 \\
-1 & 0 \\
\end{pmatrix}$ & 4 & 3\\
$III^*$ & 1728 & $\begin{pmatrix}
0 & -1 \\
1 & 0 \\
\end{pmatrix}$ & 4 & 9\\ \hline
$IV$ & 0 & $\begin{pmatrix}
0 & 1 \\
-1 & -1 \\
\end{pmatrix}$ & 3 & 4\\
$IV^*$ & 0 & $\begin{pmatrix}
-1 & -1 \\
1 & 0 \\
\end{pmatrix}$ & 3 & 8\\ \hline
\end{tabular}
\caption{\label{table monodromy in 3.3.1}J-invariants, monodromies, and associated numbers of 7-branes for fiber types.}
\end{table}
\endgroup
\par In Section \ref{ssec:3.4}, we use the results in Table \ref{table monodromy in 3.3.1} to check the consistency of the gauge symmetries on the 7-branes in F-theory compactifications on double covers times K3, which we obtained in Section \ref{ssec:3.1}. 

\subsubsection{Anomaly cancellation condition}
The cancellation condition of the tadpole without a flux determines the form of the discriminant locus in the base $\P^1 \times$ K3 of the F-theory compactification on K3 $\times$ K3. The form of the discriminant locus is as follows:
\begin{equation}
\{{\rm 24 \hspace{1.5mm} points \hspace{1.5mm} (counted \hspace{1.5mm} with \hspace{1.5mm} multiplicity)}\} \times {\rm K3}.
\end{equation}
Here, the points are counted with multiplicity assigned; the actual number of points can be smaller than 24. Thus, there are 24 7-branes, and they wrap K3 surfaces in the base. See, for example, \cite{K} for discussion. Using this, we check the consistency of the gauge symmetries in Section \ref{ssec:3.4}. 

\subsection{Consistency check of gauge symmetries}
\label{ssec:3.4}
We check the consistency of the gauge symmetries on the 7-branes obtained in Sections \ref{ssec:3.1} and \ref{ssec:3.2}, by considering monodromies around the singular fibers and the consistency condition from the anomaly. 

\subsubsection{Double Covers of $\P^1\times\P^1$ Ramified over Bidegree (4,4) Curve times K3}
As we stated in Section \ref{ssec 2.1}, the singular fibers of double covers of $\P^1\times\P^1$ (\ref{eq:2}) have j-invariant 1728. From Table \ref{table monodromy in 3.3.1} in Section \ref{ssec:3.3}, we find that the types of the singular fibers that have j-invariant 1728 are: $III$, $I^*_0$ and $III^*$. (j-invariant of type $I^*_0$ fiber can be 1728.) Corresponding gauge symmetries on the 7-branes are: $SU(2)$, $SO(8)$ and $E_7$. This agrees with the results that we obtained in Section \ref{ssec:3.1}. Monodromies of order 2 and 4 characterize the types of singular fibers, and the corresponding gauge groups on the 7-branes.
\par We saw in Section \ref{ssec:3.1} that F-theory compactifications on double covers (\ref{eq:2}) times a K3 generically have 8 type $III$ fibers. Sum of the number of 7-branes associated with 8 type $III$ fibers is 24; therefore, we confirm that this result is in agreement with the consistency condition from the cancellation of the tadpole. 
\par We saw in Section \ref{ssec:3.1} that, when two type $III$ fibers collide, they are enhanced to type $I^*_0$ fiber. From Table \ref{table monodromy in 3.3.1}, we confirm that this is consistent with the associated numbers of the 7-branes. When a triplet of type $III$ fibers coincide, they are enhanced to type $III^*$ fiber. We confirm that this is also consistent with the associated numbers of the 7-branes. Therefore, all the gauge symmetries on the 7-branes obtained in Section \ref{ssec:3.1} are in agreement with the anomaly cancellation condition. 

\subsubsection{Fermat Quartic times K3}
As we saw in Section \ref{ssec:3.2}, F-theory compactification on Fermat quartic (\ref{complete intersection in 3.2}) times a K3 has six $I_4$ fibers. From Table \ref{table monodromy in 3.3.1}, we confirm that the sum of the number of the associated 7-branes is 24. Therefore, the gauge symmetry obtained in Section \ref{ssec:3.2} is consistent with the anomaly cancellation condition.

\subsection{Models without $U(1)$ symmetry}
\label{ssec:3.5}
\subsubsection{Double covers of $\P^1\times\P^1$ ramified over bidegree (4,4) curve times K3}
We saw in Section \ref{ssec:3.1} that the most enhanced gauge group on the 7-branes in F-theory compactifications on double covers (\ref{double cover no expansion in 3.1}) times K3 is $E_7\times E_7 \times SO(8)$; for this case, double cover of $\P^1\times\P^1$ has two type $III^*$ fibers and one type $I^*_0$ fiber. We stated in Section \ref{ssec:3.1} that, for example, such double cover is given by the following equation: 
\begin{equation}
\tau^2=(t-\alpha_1)^3(t-\alpha_2)x^4+(t-\alpha_2)(t-\alpha_3)^3.
\label{most enhanced in 3.5.1}
\end{equation} 
Double cover (\ref{most enhanced in 3.5.1}) is an attractive K3 surface. Its Jacobian fibration is given by:
\begin{equation}
\tau^2=\frac{1}{4}x^3-(t-\alpha_1)^3(t-\alpha_2)^2(t-\alpha_3)^3\, x.
\label{jacobian in 3.5.1}
\end{equation}
The sum of the ranks of the reducible fibers of the Jacobian (\ref{jacobian in 3.5.1}) is 18; therefore, we conclude that its Mordell--Weil rank is 0. Jacobian fibration (\ref{jacobian in 3.5.1}) is an attractive K3 surface with a section, with a Mordell--Weil rank 0. Such a K3 surface is called an {\it extremal K3 surface}. Extremal K3 surfaces were classified in \cite{SZ}. 
\par Jacobian fibration (\ref{jacobian in 3.5.1}) has two type $III^*$ fibers and one $I^*_0$ fiber, same as double cover (\ref{most enhanced in 3.5.1}). Therefore, the reducible fiber type of Jacobian (\ref{jacobian in 3.5.1}) is $E^2_7D_4$. From Table 2 of \cite{SZ}, we find that the extremal K3 surface with reducible fiber type $E^2_7D_4$ is uniquely determined, and its transcendental lattice has the intersection matrix $\begin{pmatrix}
2 & 0 \\
0 & 2 \\
\end{pmatrix}$. Therefore, the transcendental lattice of Jacobian (\ref{jacobian in 3.5.1}) has the intersection matrix $\begin{pmatrix}
2 & 0 \\
0 & 2 \\
\end{pmatrix}$. We denote Jacobian (\ref{jacobian in 3.5.1}) by $S_{[2 \hspace{1mm} 0 \hspace{1mm} 2]}$. The Mordell--Weil group of Jacobian (\ref{jacobian in 3.5.1}) is $\Z_2$ \cite{SZ}. Therefore, we find that the global structure of the non-Abelian gauge symmetry is 
\begin{equation}
E_7^2\times SO(8) / \Z_2.
\end{equation} 
\par We conclude from the argument above that the F-theory compactification on double cover (\ref{most enhanced in 3.5.1}) times K3 does not have a $U(1)$ symmetry. The above-mentioned argument applies to the other cases, in which gauge group $E_7\times E_7\times SO(8)$ arises on the 7-branes in F-theory compactifications on double covers of $\P^1\times\P^1$ times K3. 

\subsubsection{Fermat quartic times K3}
We saw in Section \ref{ssec:3.2} that the Jacobian fibration of Fermat quartic (\ref{complete intersection in 3.2}) is the attractive K3 surface $S_{[4 \hspace{1mm} 0 \hspace{1mm} 4]}$, whose transcendental lattice has the intersection matrix $\begin{pmatrix}
4 & 0 \\
0 & 4 \\
\end{pmatrix}$. When the attractive K3 surface $S_{[4 \hspace{1mm} 0 \hspace{1mm} 4]}$ has six $I_4$ fibers, the reducible fiber type is $A^6_3$; therefore, the sum of the ranks of the reducible fibers is 18. From this, we deduce that the Mordell--Weil rank is 0, and the Jacobian $S_{[4 \hspace{1mm} 0 \hspace{1mm} 4]}$ with six $I_4$ fibers is an extremal K3 surface. The Mordell--Weil group of Jacobian $S_{[4 \hspace{1mm} 0 \hspace{1mm} 4]}$ with six $I_4$ fibers is $\Z_4\times\Z_4$ \cite{PS-S, Nish, SZ}. Therefore, we find that the global structure of the non-Abelian gauge symmetry is 
\begin{equation}
SU(4)^6 / \Z_4\times\Z_4.
\end{equation}  
\par We conclude that the F-theory compactification on Fermat quartic (\ref{complete intersection in 3.2}) times K3 does not have a $U(1)$ gauge symmetry.

\section{Matter fields on 7-branes}
\label{sec:4}
In this section, we compute the potential matter spectra that arise on the 7-branes in F-theory compactifications on double covers times K3, and compactifications on the Fermat quartic times K3. Matter fields on the 7-branes arise from rank one enhancements of singularities. F-theory compactifications on product K3 $\times$ K3 gives a four-dimensional theory with $N=2$ supersymmetry. 7-branes in F-theory on direct product K3 $\times$ K3 are parallel; therefore, the only light matter fields on the 7-branes without a flux are the adjoints of the gauge groups. With fluxes, half the supersymmetry is broken, and F-theory on K3 $\times$ K3 with a flux gives four-dimensional theory with $N=1$ supersymmetry. By including fluxes, hypermultiplets split into vector-like pairs. These vector-like pairs are candidates for the matter spectra with a flux; vector-like pairs may vanish owing to the tadpole. 
\par For special cases such as when K3 is attractive, the complex structure of K3 can be determined. This enables us to study the cancellation of the tadpole in detail. Fermat quartic is the attractive K3 surface $S_{[8 \hspace{1mm} 0 \hspace{1mm} 8]}$. We will find in Section \ref{ssec:4.1} that the tadpole can be cancelled for F-theory compactifications on the Fermat quartic times some appropriate attractive K3 surface, by including sufficiently many 3-branes. Vector-like pairs actually arise for this particular compactification.

\subsection{Cancellation of tadpole for Fermat quartic with flux}
\label{ssec:4.1}
The complex structure of an attractive K3 surface is specified by its transcendental lattice \cite{SI}. See, for example, \cite{K} for a review of this correspondence. For an attractive K3 surface, the intersection matrix of its transcendental lattice is a 2 $\times$ 2 integral matrix of the following form:
\begin{equation}
\begin{pmatrix}
2a & b \\
b & 2c \\
\end{pmatrix},
\end{equation}
where $a$,$b$,$c$ are in $\Z$. In this note, we denote the attractive K3 surface, whose transcendental lattice has the intersection matrix $\begin{pmatrix}
2a & b \\
b & 2c \\
\end{pmatrix}$, by $S_{{\rm [}2a \hspace{1mm} b \hspace{1mm} 2c{\rm]}}$. 
\par In the presence of 4-form flux $G$, the tadpole cancellation condition \cite{VW, SVW} for F-theory on product K3$\times$K3 is given by: 
\begin{equation}
\frac{1}{2}\int_{{\rm K3}\times{\rm K3}} G\wedge G+N_3=\frac{1}{24}\chi({\rm K3}\times {\rm K3})=24,
\label{eq:tac}
\end{equation} 
where $N_3$ represents the number of 3-branes turned on. The 4-form flux $G$ is subject to the following quantization condition \cite{W}: 
\begin{equation}
G\in H^4({\rm K3}\times {\rm K3},\Z).
\end{equation} 
\par \cite{AK} discussed M-theory flux compactification on the product of attractive K3's $S_1\times S_2$. \cite{AK} considered the following decomposition of the 4-form flux $G$:
\begin{equation}
G=G_0+G_1,
\end{equation}
where
\begin{eqnarray}
G_0 & \in & H^{1,1}(S_1,\R)\otimes H^{1,1}(S_2,\R) \\
G_1 & \in & H^{2,0}(S_1,\C) \otimes H^{0,2}(S_2,\C)+{\rm h.c.}
\end{eqnarray}
Under the constraints  
\begin{equation}
G_0=0
\end{equation}
and
\begin{equation}
N_3=0,
\label{eq:flux assump}
\end{equation} 
all the pairs of attractive K3 surfaces $S_{[2a \hspace{1mm} b \hspace{1mm} 2c]}$ $\times$ $S_{[2d \hspace{1mm} e \hspace{1mm} 2f]}$, for which tadpole cancellation condition (\ref{eq:tac}) can be satisfied, were determined in \cite{AK}.
\cite{BKW} relaxed the condition (\ref{eq:flux assump}) to
\begin{equation}
N_3\ge 0
\end{equation}
and extended the list of attractive K3 pairs, for which the tadpole is cancelled. There are only finitely many attractive K3 pairs in both the lists in \cite{AK} and \cite{BKW}; therefore, the complex structure moduli are fixed in \cite{AK,BKW}.
\par Fermat quartic surface $S_{[8 \hspace{1mm} 0 \hspace{1mm} 8]}$ appears in the list of \cite{BKW}\footnote{\cite{BKW} uses a different notational convention for attractive K3 surfaces. They denote the subscript by [$a$ $b$ $c$] for attractive K3, whose transcendental lattice has the intersection matrix $\begin{pmatrix}
2a & b \\
b & 2c \\
\end{pmatrix}$. In the list of \cite{BKW}, [1 0 1] represents the attractive K3 surface, which is denoted by $S_{[2 \hspace{1mm} 0 \hspace{1mm} 2]}$ in this note. Similarly, [4 0 4] in the list of \cite{BKW} represents the Fermat quartic surface $S_{[8 \hspace{1mm} 0 \hspace{1mm} 8]}$.}. There is only 1 pair in the list, which contains $S_{[8 \hspace{1mm} 0 \hspace{1mm} 8]}$: $S_{[8 \hspace{1mm} 0 \hspace{1mm} 8]}\times S_{[2 \hspace{1mm} 0 \hspace{1mm} 2]}$. For this attractive K3 pair, the tadpole anomaly is cancelled by turning on sufficiently many 3-branes. ($N_3=8$ for the pair $S_{[8 \hspace{1mm} 0 \hspace{1mm} 8]}\times S_{[2 \hspace{1mm} 0 \hspace{1mm} 2]}$\cite{BKW}.) For this pair, we can say that vector-like pairs will arise. $SU(4)^6$ gauge group arises on the 7-branes in the F-theory compactification on the Fermat quartic $S_{[8 \hspace{1mm} 0 \hspace{1mm} 8]}$ times K3, and therefore matter fields arise from $A_3$ singularities. 

\subsection{Matter spectra}
 
\subsubsection{Matter spectra on double cover of $\P^1\times\P^1$ ramified over a bidegree (4,4) curve times K3}
We deduced in Section \ref{ssec:3.1}, that the gauge groups on the 7-branes in F-theory on double covers (\ref{double cover no expansion in 3.1}) times K3 are $SU(2)$, $SO(8)$, or $E_7$. Therefore, the corresponding singularity structures are $A_1$, $D_4$, or $E_7$. 
\par Matter fields do not arise from an $A_1$ singularity. As we saw in Section \ref{ssec:3.1}, $SU(2)^8$ gauge group generically arises on the 7-branes in F-theory on a double cover of $\P^1\times\P^1$ ramified over a bidegree (4,4) curve times K3. Therefore, matter fields do not arise for this generic configuration of the gauge groups. Matter fields arise only when singular fibers collide, and the gauge group on the 7-branes is enhanced.   
\par When two type $III$ fibers collide, the resulting enhanced fiber is a $I^*_0$ fiber. When a triplet of type $III$ fibers coincide, the resulting fiber is a $III^*$ fiber. These enhanced singular fibers correspond to $D_4$ and $E_7$ singularities, respectively. Therefore, matters arise from $E_7$ and $D_4$ singularities.
\par There are two enhancements for the $E_7$ singularity: 
\begin{eqnarray}
E_6 & \subset & E_7 \\
A_6 & \subset & E_7.
\end{eqnarray}
The adjoint of $E_7$ decomposes into irreducible representations of $E_6$ as \cite{Sla}
\begin{equation}
{\bf 133}={\bf 78}+{\bf 27}+\overline{\bf 27}+{\bf 1}.
\end{equation}
Similarly, the adjoint of $E_7$ decomposes into irreducible representations of $A_6$ as
\begin{equation}
{\bf 133}={\bf 48}+[ \, {\bf 7}(\, \ytableausetup{boxsize=.6em}\ytableausetup
{aligntableaux=center}\begin{ytableau}
 \\
\end{ytableau} \,)+\overline{\bf 7}+{\bf 35}(\, \ytableausetup{boxsize=.6em}\ytableausetup
{aligntableaux=center}\begin{ytableau}
 \\
 \\
 \\
\end{ytableau} \,)+\overline{\bf 35} \, ]+{\bf 1}.
\end{equation}
Therefore, we see that the matters arising on the 7-branes from the $E_7$ singularity are the adjoints ${\bf 78}$ of $E_6$ and the adjoints ${\bf 48}$ of $A_6$ without fluxes. By including a flux, vector-like pairs ${\bf 27}+\overline{\bf 27}$ of $E_6$ and ${\bf 35}+\overline{\bf 35}$ and ${\bf 7}+\overline{\bf 7}$ of $A_6$ could also arise. 
\par The enhancement for the $D_4$ singularity is 
\begin{equation}
A_3\subset D_4.
\end{equation}
Under this enhancement, the adjoint of $D_4$ decomposes into irreducible representations of $A_3$ as 
\begin{equation}
{\bf 28}={\bf 15}+{\bf 6}(\, \ytableausetup{boxsize=.6em}\ytableausetup
{aligntableaux=center}\begin{ytableau}
 \\
 \\
\end{ytableau} \,)+\overline{\bf 6}+{\bf 1}.
\end{equation}
It follows that the adjoints ${\bf 15}$ of $A_3$ will arise on the 7-branes from the $D_4$ singularity. Vector-like pairs ${\bf 6}+\overline{\bf 6}$ could also arise with a flux. This enhancement is also discussed in \cite{MTmatter}. 

\subsubsection{Matter spectra on Fermat Quartic times K3}
As we saw in Section \ref{ssec:3.2}, the Fermat quartic surface, presented as complete intersection (\ref{complete intersection in 3.2}), has six $I_4$ fibers. $I_4$ fiber corresponds to an $A_3$ singularity. Therefore, in F-theory compactification on the Fermat quartic times a K3, matter fields arise from $A_3$ singularities. 
\par The enhancement for the $A_3$ singularity is
\begin{equation}
A_2\subset A_3.
\end{equation} 
Under this enhancement, the adjoint of $A_3$ decomposes into irreducible representations of $A_2$ as
\begin{equation}
{\bf 15}={\bf 8}+{\bf 3}(\, \ytableausetup{boxsize=.6em}\ytableausetup
{aligntableaux=center}\begin{ytableau}
 \
\end{ytableau} \,)+\overline{\bf 3}+{\bf 1}.
\end{equation}
It follows that the adjoints ${\bf 8}$ of $A_2$ will arise on the 7-branes. The vector-like pairs ${\bf 3}+\overline{\bf 3}$ could also arise by including flux.
\par We saw in Subsection \ref{ssec:4.1} that the tadpole is cancelled for F-theory compactification on the pair $S_{[8 \hspace{1mm} 0 \hspace{1mm} 8]}\times S_{[2 \hspace{1mm} 0 \hspace{1mm} 2]}$ by including sufficiently many 3-branes. Therefore, in F-theory compactification on $S_{[8 \hspace{1mm} 0 \hspace{1mm} 8]}\times S_{[2 \hspace{1mm} 0 \hspace{1mm} 2]}$ with flux, the vector-like pairs ${\bf 3}+\overline{\bf 3}$ also arise on the 7-branes.

\section{Conclusion}
\label{sec:5}
Double covers of $\P^1\times\P^1$ ramified along a bidegree (4,4) curve and complete intersections of two bidegree (1,2) hypersurfaces in $\P^1\times\P^3$ are genus-one fibered K3 surfaces. Generic members of these K3 surfaces do not have a global section. In this note, we investigated the gauge symmetries and matter fields that arise on the 7-branes in F-theory compactifications on these K3 genus-one fibrations without a section times K3. To study the gauge theories and matter spectra in detail, we particularly focused on the double covers of $\P^1\times\P^1$ given by the following equation:
\begin{equation}
\tau^2=a_1(t)x^4+a_2(t).
\label{eq:last}
\end{equation}
($a_1(t)$ and $a_2(t)$ are polynomials in $t$ of the highest degree of 4.)
Genus-one fibers of double covers (\ref{eq:last}) possess complex multiplication of order 4; this property enables us to analyze the non-Abelian gauge symmetries that arise on the 7-branes in detail. For complete intersection K3 surface, we particularly focused on the complete intersection given by the intersection of the following two hypersurfaces in $\P^1\times\P^3$:
\begin{eqnarray}
\label{array last}
x^2_1+x^2_3+2tx_2x_4 & = & 0 \\
x^2_2+x^2_4+2tx_1x_3 & = & 0 \nonumber
\end{eqnarray}
We saw in Section \ref{ssec:2.2} that complete intersection (\ref{array last}) is isomorphic to the Fermat quartic surface in $\P^3$:
\begin{equation}
x^4+y^4+z^4+w^4=0.
\label{Fermat last}
\end{equation} 
\par Generic members of double covers (\ref{eq:last}) have eight type $III$ fibers; therefore, a $SU(2)^8$ gauge group arises on the 7-branes in F-theory compactifications on double covers (\ref{eq:last}) times K3. When type $III$ fibers collide, $SU(2)^2$ gauge symmetry on the 7-brane is enhanced to the $SO(8)$ gauge group. When a triplet of type $III$ fibers coincide, $SU(2)^3$ gauge symmetry is enhanced to $E_7$ gauge symmetry. Further enhancement breaks the Calabi--Yau condition. The most enhanced gauge symmetry on the 7-branes in F-theory on the double cover of $\P^1\times\P^1$ ramified over a bidegree (4,4) curve times K3 is $E_7\times E_7\times SO(8)$. When the non-Abelian gauge symmetry is enhanced to $E_7\times E_7\times SO(8)$, the Jacobian fibration of the double cover has the Mordell--Weil group of rank 0; therefore, for such cases, F-theory compactifications on double covers times K3 do not have a $U(1)$ gauge field. 
\par Fermat quartic, presented as complete intersection (\ref{array last}), has six $I_4$ fibers; $SU(4)^6$ gauge group arises on the 7-branes in F-theory compactifications on Fermat quartic (\ref{array last}) times K3. The Jacobian fibration of the Fermat quartic with six $I_4$ fibers has the Mordell--Weil group of rank 0; therefore, F-theory compactifications on Fermat quartic (\ref{array last}) times K3 do not have a $U(1)$ gauge symmetry. 
\par All 7-branes are parallel in the F-theory compactification on direct product K3 $\times$ K3; therefore, the only light matter fields on the 7-branes are the adjoints of the gauge groups without a flux. The included flux breaks half of $N=2$ supersymmetry, and vector-like pairs can also arise. We showed that the tadpole can be cancelled for the F-theory compactification on the product of the Fermat quartic $S_{[8 \hspace{1mm} 0 \hspace{1mm} 8]}$ times the attractive K3 surface $S_{[2 \hspace{1mm} 0 \hspace{1mm} 2]}$. As a result, we confirmed that vector-like pairs arise in the flux compactification on $S_{[8 \hspace{1mm} 0 \hspace{1mm} 8]}\times S_{[2 \hspace{1mm} 0 \hspace{1mm} 2]}$.      

\section*{Acknowledgments}

We would like to thank Shigeru Mukai for discussions. This work is supported by Grant-in-Aid for JSPS Fellows No. 26$\cdot$2616.

\end{document}